\documentclass[12pt,fleqn]{article}

\usepackage[english,bulgarian]{babel}

\usepackage[cp1251]{inputenc}

\usepackage{amsmath,amssymb}
\usepackage{indentfirst}

\usepackage[dvips]{graphicx}

\author{Vassil K. Tinchev}

\title{Measuring of the Compact Objects' Parameters by Analysis of their Shadow}
\author{Vassil K. Tinchev\thanks{E-mail: tintschev@gmail.com}\\
{\footnotesize  Department by Physics, Faculty of Hydraulic Engineering,}\\
{\footnotesize  University of Architecture, Civil Engineering and Geodesy,}\\
{\footnotesize  1 Hristo Smirnenski Blvd., 1164 Sofia, Bulgaria; }\\
{\footnotesize  Department of Differential Equations and Mathematical Physics,}\\
{\footnotesize  Institute of Mathematic and Informatics, Bulgarian Academy of Sciences,}\\
{\footnotesize  Acad. Georgi Bonchev Str., Block 8, 1113 Sofia, Bulgaria; }\\
{\footnotesize  Sofia High School of Mathematics,}\\
{\footnotesize  61 Iskar Str., 1000 Sofia, Bulgaria }\\}

\begin{document}

\maketitle

\begin{abstract}
The following work presents a sufficiently general method for finding the parameters that characterise self-gravitating compact objects when their shadow contour is explicitly set.
This method can be used in various algorithms to analyse experimental data for the shadow contour of specific compact objects.\\

\end{abstract}

\section*{\begin{center}Introduction\end{center}}

Известно е, че контурът на сянката на компактните самогравитиращи обекти зависи от метриката на пространство-времето в тяхната околност и съответно съдържа информация за тях \cite{FalckeEtAll}, \cite{Luminet}. Сенките на черните дупки на Kerr-Newmann са изследвани в \cite{Bardeen}-\cite{bibl2}, като първите резултати са намерени в края на миналия век. В последните години се водят интензивни изследвания в това направление \cite{AmarillaEiroaGiribet}-\cite{s45} и вече има немалко работи в които за конкретни компактни самогравитиращи обекти са получени в експлицитен вид параметрични уравнения, задаващи контура на тяхната сянка в т.нар. небесни координати \cite{bibl1}. През тази година бяха представени първите експериментални резултати на колаборацията \textit{Event Horizon Telescope} \cite{EventHorizonI}-\cite{EventHorizonVI} за сянката на свръхмасивната черна дупка в центъра на галактиката M87. Очаква се да бъдат получени данни за контура на сянката и на други централногалактични свръхмасивни черни дупки, включително и за тази в центъра на Нашата Галактика \textit{Sagittarius A*}. Възниква въпросът какво можем да кажем за природата на един компактен самогравитиращ обект, когато наблюдаваме контура на неговата сянка? Или по-конкретно, ако предположим, че наблюдаваме сянката на определен тип компактен обект, то как да преценим доколко това предположение е уместно и съответно да оценим параметрите, характеризиращи дадения обект и положението на наблюдателя спрямо него? При всеки стационарен източник на гравитационно поле имаме поне три такива параметъра. Това са масата на обекта $M$, ъгловият момент на единица маса $a$ и ъгълът $\theta_{0}$ под който наблюдателя вижда оста на въртене на обекта - т.нар ъгъл на инклинация. Съответно контурът на сянката на обекта зависи от тези параметри. Най-простият случай е този на въртяща се черна дупка в Айнщайновата теория на гравитацията. Незаредената черна дупка се описва от решението на Kerr \cite{Kerr}, а тази с електричен заряд от решението на Kerr-Newman \cite{Newman1}, \cite{Newman2}. Контурът на сянката на незаредената черна дупка зависи само от тези три параметъра. Ако стационарната черна дупка има електричен заряд $Q$, то неговата стойност ще бъде допълнителен четвърти параметър от който зависи контура на сянката й. Ясно е, че контурите на сянката на компактните обекти от един и същи тип (например черни дупки на Kerr) ще се представят чрез даден клас криви, като отделните екземпляри на класа ще се характеризират с параметрите в решението на уравненията на гравитационното поле, задаващо определената от компактният обект метрика на пространство-времето. При стационарните обекти допълнителен параметър от който зависи контура на тяхната сянка, е ъгълът на инклинация $\theta_{0}$. С други думи, ако означим с $\alpha$ и $\beta$ "небесните координати"\ на контура на сянката, то те ще бъдат функции на параметрите $M$, $a$, $Q$ и т.н. от които зависи определената от компактният обект метрика на пространство-времето, а също така и от ъгъла на инклинация $\theta_{0}$ - $\alpha\equiv\alpha(M,a,...,\theta_{0},r)$, $\beta\equiv\beta(M,a,...,\theta_{0},r)$. Тук $r\in\left[r_{min},r_{max}\right]$ е аргументът чрез който се параметризира контура на сянката. Изобщо казано в експлицитните изрази за $\alpha$ и $\beta$, аргументът $r$ задава радиусите на нестабилни кръгови фотонни орбити. Когато наблюдаваме сянката на компактният обект, то отговорът на въпроса какво можем да кажем за неговата природата, се свежда до една конкретна задача. Трябва да укажем метод с помощта на който да определяме параметрите $M$, $a$, ... и $\theta_{0}$, характеризиращи компактните самогравитиращи обекти, когато имаме само графиката на контура на тяхната сянка и можем да правим измервания на отсечки и ъгли върху тази графика. В настоящата работа е представен един такъв метод за определяне параметрите на компактните самогравитиращи обекти чрез измервания върху контура на сянката им. Той може да се използва в различни алгоритми за анализ на експерименталните данни за контура на сянката на конкретни компактни обекти. Същността на метода е илюстрирана за случая на черна дупка на Kerr, като е указано как следва да се постъпи във всяка друга подобна ситуация при която контурът на сянката на компактния обект се задава експлицитно.
\newpage
\section*{\begin{center}Case of the Kerr black hole\end{center}}

Контурът на сянката на една Керовска черна дупка се параметризира по следния начин:
\begin{equation}\label{solalphabetaA2}
    \begin{array}{ll}
    \alpha=\frac{1}{a(r-M)\sin\theta_{0}}\left[r^2(r-3M) + a^2(r+M)\right],\\
    \\
    \beta=\pm\sqrt{\frac{2\left[r^{2}(r^2-3M^2)+a^2(r^2+M^2)\right]}{(r-M)^{2}}-a^{2}\sin^{2}\theta_{0}-
      \alpha^{2}}.
    \end{array}
\end{equation}
Тук $\alpha(r)$ и $\beta(r)$ са небесните координати на контура на сянката, $M$, $a$ и $\theta_{0}$ са съответно масата на черната дупка, нейния ъглов момент на единица маса и ъгъла под който наблюдателя вижда оста на въртене на черната дупка - ъгъла на инклинация. Аргументът $r$ чрез който се параметризира контура на сянката, както вече казахме, задава радиусите на т.нар. критични кръгови фотонни орбити. При наблюдение на сянката на черната дупка може да определим координатата $\beta_{i}\equiv\beta(M,a,s,r_{i})$ на всяка една точка $C_{i}=(\alpha_{i},\beta_{i})$ от контура непосредствено - в дадени избрани от нас единици $M_{u}$, съответстващи на мащаба в който е начертан контурът на сянката на черната дупка. Така е защото оста $\alpha$ е ос на симетрия за сянката. По тази причина по-натам ще разглеждаме само "горната"\ половина на сянката - за която непосредствено измеряемата координата $\beta$ е по-голяма от нула. Координата $\alpha_{i}\equiv\alpha(M,a,s,r_{i})$ на огледално разположените (спрямо оста $\alpha$) точки $C_{i}(\alpha_{i},\beta_{i})$ и $\widetilde{C}_{i}(\alpha_{i},-\beta_{i})$ от "горната"\ и "долната"\ половина на сянката, може да се определят с точност до произволна транслация по оста $\alpha$ на началото на координатната система. С други думи наблюдението на сянката в "хоризонтално"\ направление (т.е. перпендикулярно на $\beta$), позволява да се измери непосредствено в единици $M_{u}$ не $\alpha_{i}$ координатата, а разликата $\Delta\alpha_{ij}\equiv\alpha_{j}-\alpha_{i}$, определена за всеки две точки $C_{i}=(\alpha_{i},\beta_{i})$ и $C_{j}=(\alpha_{j},\beta_{j})$ от контура. Удобно е да използваме и друга непосредствено измеряема по контура на сянката величина, която се определя чрез производните по $r$ на небесните координати $\alpha(r)$ и $\beta(r)$. Такава величина е тангенсът на ъгъла $\varsigma$, който сключва с оста $\alpha$ допирателният към контура на сянката вектор в произволна точка $C=(\alpha,\beta)$. За него имаме:
\begin{equation}\label{defT}
  \tan\varsigma=\frac{d\beta}{d\alpha}=\frac{\beta'}{\alpha'}\equiv T;\ \alpha'\equiv\frac{d\alpha}{dr},\ \beta'\equiv\frac{d\beta}{dr}.
\end{equation}
Съласно (\ref{solalphabetaA2}):
\begin{equation}\label{Trbetaalpha}
  T=\frac{-r^{2}(r-3M)-a^{2}(r+M)(1-s^{2})}{\beta as(r-M)}=-\frac{\alpha}{\beta}+\frac{as(r+M)}{\beta(r-M)},\ s\equiv\sin\theta_{0}.
\end{equation}

Нека представим (\ref{solalphabetaA2}) и (\ref{Trbetaalpha}) чрез съответните им полиномиални уравнения. Ясно е, че при $r\neq M$, $a\neq 0$, $s\neq 0$ и $\beta\neq 0$ това винаги е възможно. Имаме:
\begin{equation}\label{PolysysalphabetaT}
    \begin{array}{ll}
    P_{\alpha}\equiv as\alpha(r-M)-r^2(r-3M)-a^2(r+M)=0,\\
    \\
    P_{\beta}\equiv(\beta^{2}+a^{2}s^{2}+\alpha^{2})(r-M)^{2}-2\left[r^{2}(r^2-3M^2)+a^2(r^2+M^2)\right]=0,\\
    \\
    P_{T}\equiv(\beta T+\alpha)(r-M)-as(r+M)=0.
    \end{array}
\end{equation}
Можем да заменим полиномът $P_{T}$ с полином $P_{U}$ ($U\equiv 1/T$). Съгласно (\ref{Trbetaalpha}) е изпълнено:
\begin{equation}\label{PolyU}
  P_{U}\equiv U\left[r^{2}(r-3M)+a^{2}(r+M)(1-s^{2})\right]+\beta as(r-M)=0.
\end{equation}

Задачата, която следва да решим, е намирането с помощта на (\ref{PolysysalphabetaT}), на поне едно ново полиномиално уравнение, чиито аргументи са параметрите на Керовската черна дупка ($M$ и $a$), параметъра характеризиращ положението на наблюдателя спрямо оста на въртене на черната дупка ($s$), заедно с величините, които можем да измерваме непосредствено, наблюдавайки контура на сянката ($\beta$ и $T$). Очевидно за постигането на тези цел трябва в системата (\ref{PolysysalphabetaT}) да се елиминират едновременно $\alpha$ и $r$. Необходимо да изследваме полиномиалната система (\ref{PolysysalphabetaT}) с помощта на метода, използващ базиси на Gr\"{o}bner \cite{Groebner}. Оказва се, че алгоритъмът на Buchberger \cite{Buchberger}, приложен към (\ref{PolysysalphabetaT}), е напълно изчислим и води до следния експлицитен резултат:
\begin{equation}\label{solP}
    \begin{array}{ll}
    P\equiv 64a^{6}M^{2}(1-s^{2})^{3}+48\beta^{2}a^{4}M^{2}(1-s^{2})\left[(2+T^{2})s^{2}-2(1+T^{2})\right]-\\
    \\
    \ \ \ \ -16\beta^{3}T^{3}a^{3}M^{2}s^{3}-\beta^{4}(1+T^{2})^{2}\left[27M^{4}-a^{4}(1-s^{2})^{2}\right]-\\
    \\
    \ \ \ \ -6\beta^{4}(1+T^{2})a^{2}M^{2}\left[(5+T^{2})s^{2}-5(1+T^{2})\right]+\\
    \\
    \ \ \ \ +\beta^{6}(1+T^{2})^{2}\left[(1+T^{2})(M^{2}-a^{2})+a^{2}s^{2}\right]=0.
    \end{array}
\end{equation}
Виждаме, че ако вземем три точки от "горната"\ половина на наблюдаваната от нас сянка на черната дупка, за които $(\beta,T)$ са съответно $(\beta_{1},T_{1})$, $(\beta_{2},T_{2})$ и $(\beta_{3},T_{3})$, то когато $(\beta_{1},T_{1})\neq (\beta_{2},T_{2})\neq (\beta_{3},T_{3})$, е възможно по принцип да определим $M$, $a$ и $s$, решавайки следната полиномиална система уравнения:
\begin{equation}\label{sysP}
    \begin{array}{ll}
    P(\beta_{1},T_{1},M,a,s)=0,\\
    P(\beta_{2},T_{2},M,a,s)=0,\\
    P(\beta_{3},T_{3},M,a,s)=0
    \end{array}
\end{equation}
Ясно е, че при такова едно разглеждане на (\ref{solP}) само по три точки от контура на сянката, изобщо казано няма как да получим едно единствено решение. Освен това едва ли ще е възможно да се реши системата точно, имайки предвид степените на мономите по $M$, $a$ и $s$. Разбира се винаги може да решим числено въпросната полиномиална система уравнения за $M$, $a$ и $s$. Освен това ние имаме и свободата да изберем повече от три точки от контура на сянката, т.е. да търсим числени решения на (\ref{sysP}) относно $M$, $a$ и $s$, взимайки колкото искаме тройки стоиности $\beta_{i}$ и $T_{i}$. По този начин ще имаме възможност да определим ефективно истинските стойности на $M$, $a$ и $s$, тъй като те ще присъстват сред решенията на (\ref{sysP}) за всяка от избраните тройки точки от контура на сянката на нашата черна дупка. Всъщност свободата да вземем колкото искаме точки от контура на сянката, подсказва един възможен изход, позволяващ да се реши напълно експлицитно до край задачата за намиране на ъгъла на инклинация и параметрите на въртящата се черна дупка, имайки данните за контура на нейната сянка. Полиномиалното уравнение (\ref{sysP}) има следната структура:
\begin{equation}\label{eqPStructure}
  P(\beta,T,M,a,...,s)=\sum_{j=1}^{K}A_{j}(\beta,T)p_{j}(M,a,...,s),
\end{equation}
Ако разделим това уравнение на конкретен полином $p_{c}(M,a,...,s)\neq 0$, ще получим:
\begin{equation}\label{eqPLinPres}
  \frac{P(\beta,T,M,a,...,s)}{p_{c}(M,a,...,s)}=-B(\beta,T)+\sum_{j=1,\ j\neq c}^{K}A_{j}(\beta,T)q_{j}(M,a,...,s),\ 1\leq c\leq K,
\end{equation}
където:
\begin{equation}\label{eqPLinPresCoefficienti}
    \begin{array}{l}
    B(\beta,T)=-A_{c}(\beta,T),\\
    \\
    q_{j}(M,a,...,s)=\dfrac{p_{j}(M,a,...,s)}{p_{c}(M,a,...,s)},\ j\neq c.
    \end{array}
\end{equation}
По-натам за удобство ще преномерираме рационалните функции $q_{j}(M,a,...,s)$ по такъв начин, че $j\in\{1,2,3,...,K-1\}$. Нека $\mathfrak{q}$ е множеството рационални функции $q_{j}(M,a,...,s)$. Избираме измежду тях подмножество $\mathfrak{Q}$, състоящо се от членове $Q_{j}(M,a,...,s)\in\mathfrak{Q}\subset\mathfrak{q}$, които са такива, че за $Q_{1}(M,a,...,s)$, $Q_{2}(M,a,...,s)$, ... и $Q_{g}(M,a,...,s)$ ($g\leq k\equiv K-1$), системата:
\begin{equation}\label{sysasmin}
    \begin{array}{l}
    Q_{1}(M,a,...,s)=v_{1},\\
    Q_{2}(M,a,...,s)=v_{2},\\
    ...\\
    Q_{g}(M,a,...,s)=v_{g},
    \end{array}
\end{equation}
е разрешима относно параметрите на черната дупка $M$, $a$, ..., и синуса от ъгъла на инклинация $s$. Тук сме отчели това, че ако разглеждаме не Керовска черна дупка, то трябва да имаме не три, а повече уравнения в минималната разрешима относно параметрите на черната дупка система рационални уравнения (\ref{sysasmin}). Това е изпълнимо условие. Един възможен избор би могъл да е този при който степените на неизвестните параметри $M$, $a$, $s$, ... във функциите $Q_{1}$, $Q_{2}$, ..., $Q_{g}$ са по-ниски от съответните степени във всяка друга функция $q_{j}\in\mathfrak{q}$. Ако системата (\ref{sysasmin}) е такава, че $k=g$, то толкова по-добре. Но за съжаление това не винаги е постижимо.\\
За да намерим как $v_{1}$, $v_{2}$, ..., $v_{g}$ се изразяват от данните за контура на сянката (в Керовския случай $g=3$), трябва да вземем $k$ различни точки от контура и да решим относно $q_{j}$ следната линейна система уравнения:
\begin{equation}\label{sysqjLin}
  \sum_{j=1}^{k} A_{j}(\beta_{i},T_{i})q_{j}=B(\beta_{i},T_{i}).
\end{equation}
Разбира се трябва така да сме подбрали точките от контура на сянката, че да не е равна на нула детерминантата на матрицата $A_{j}(\beta_{i},T_{i})$, което при положение, че имаме множество от точки с мощност континуум, не би трябвало да е трудно. От всичките $k$ решения за функциите $q_{j}$, са важни тези, които позволяват да се намерят експлицитно параметрите на черната дупка и ъгъла на инклинация - т.е. решенията за $Q_{j}\in\mathfrak{Q}\subset\mathfrak{q}$, които са три ($j\in\{1,2,3\}$) в избрания случай. Другите решения за $q_{j}$ задават връзки между различните $\beta_{i}$ и $T_{i}$ на избраните точки от контура на сянката, които са конкретни съотношения, валидни за сянката на дадения тип черна дупка. Тях можем да използваме като критерий за това дали изследваната сянка, е сянка на точно този тип черна дупка, който предполагаме - в дадения случай Керовска черна дупка. Подобен критерий е и условието $|s|<1$.\\
Преди да приложим описания алгоритъм за избрания конкретен пример на въртяща се незаредена черна дупка, нека отбележим, че при други типове черни дупки е възможно по метода на базиси на Gr\"{o}bner да се намерят не едно, а повече независими уравнения от типа на (\ref{solP}). В някои случаи може да се окаже удобно да се използва цялата съвкупоност от полиномиални уравнения, а не само това, което е от най-ниска степен. Това не пречи да се следва алгоритъма. Ще трябва рационалните функции $P$, $A_{j}$, $q_{j}$ и $B$ да номерираме с допълнителни индекси - съответно като $P_{h}$, $A_{hj}$, $q_{hj}$ и $B_{h}$, където $h$ се изменя от $1$ до броя различни уравнения от типа на (\ref{solP}) - да речем $l$.

И така, нека сега приложим всичко казано дотук относно възможностите за анализ на експлицитно намерените с помощта на базиси на Gr\"{o}bner, полиномиални уравнения от типа на (\ref{solP}), за случая с контура на сянката на въртяща се незаредена черна дупка на Kerr. Необходимо е да представим (\ref{solP}) в същия вид като (\ref{eqPLinPres}). Но преди това нека видим какви са полиномите $p_{j}(M,a,...,s)$. Съгласно (\ref{solP}) имаме:
\begin{equation}\label{pj}
    \begin{array}{l}
    p_{1}\equiv a^{2}s^{2},\ p_{2}\equiv M^{2}-a^{2},\ p_{3}\equiv a^{2}M^{2}s^{2},\ p_{4}\equiv a^{4}(1-s^{2})^{2}+30a^{2}M^{2}-27M^{4},\\
    \\
    p_{5}\equiv a^{3}M^{2}s^{3},\ p_{6}\equiv a^{4}M^{2}(1-s^{2}),\ p_{7}\equiv a^{4}M^{2}s^{2}(1-s^{2}),\ p_{8}\equiv a^{6}M^{2}(1-s^{2})^{3}.
    \end{array}
\end{equation}
Съответно полиномите $A_{j}(\beta,T)$ от (\ref{eqPStructure}) за Керовската черна дупка имат вида:
\begin{equation}\label{AKerr}
    \begin{array}{l}
    A_{1}\equiv A_{1}(\beta,T)=\beta^{6}(1+T^{2})^{2},\\
    \\
    A_{2}\equiv A_{2}(\beta,T)=\beta^{6}(1+T^{2})^{3},\\
    \\
    A_{3}\equiv A_{3}(\beta,T)=-6\beta^{4}(1+T^{2})(5+T^{2}),\\
    \\
    A_{4}\equiv A_{4}(\beta,T)=\beta^{4}(1+T^{2})^{2},\\
    \\
    A_{5}\equiv A_{5}(\beta,T)=-16\beta^{3}T^{3},\\
    \\
    A_{6}\equiv A_{6}(\beta,T)=-96\beta^{2}(1+T^{2}),\\
    \\
    A_{7}\equiv A_{7}(\beta,T)=48\beta^{2}(2+T^{2}),\\
    \\
    A_{8}\equiv A_{8}(\beta,T)=64.
    \end{array}
\end{equation}
Изборът на рационалните функции $q_{j}(M,a,.,s)$ може да се направи по различни начини. Тъй като (\ref{solP}) е валидно когато $a\neq 0$ и $s\neq 0$, то за да получим $q_{j}(M,a,.,s)$ е достатъчно да разделим на $p_{1}=a^{2}s^{2}$. Съответно списъкът с възможни рационални функции за $q_{j}$, които можем да получим от (\ref{solP}), е:
\begin{equation}\label{qj}
    \begin{array}{l}
    q_{1}\equiv \dfrac{p_{2}}{p_{1}}\ ,\ q_{2}\equiv \dfrac{p_{3}}{p_{1}}\ ,\ q_{3}\equiv \dfrac{p_{4}}{p_{1}}\ ,\ q_{4}\equiv \dfrac{p_{5}}{p_{1}}\ ,\\
    \\
    q_{5}\equiv \dfrac{p_{6}}{p_{1}}\ ,\ q_{6}\equiv \dfrac{p_{7}}{p_{1}}\ ,\ q_{7}\equiv \dfrac{p_{8}}{p_{1}}\ .
    \end{array}
\end{equation}
Виждаме, че при така избраните $q_{j}$, следва да вземем седем (т.е. $k=7$) различни точки от контура на сянката на Керовската черна дупка, за да съставим линейната система (\ref{sysqjLin}), която в дадения случай се свежда до:
\begin{equation}\label{sysqjLinKerr}
  \sum_{j=1}^{7} A_{j}(\beta_{i},T_{i})q_{j}=B(\beta_{i},T_{i})\equiv B_{i}.
\end{equation}
Тук съгласно (\ref{solP}) и (\ref{AKerr}):
\begin{equation}\label{ABKerr}
    \begin{array}{l}
    A_{i1}\equiv A_{2}(\beta_{i},T_{i})=\beta_{i}^{6}(1+T_{i}^{2})^{3},\\
    \\
    A_{i2}\equiv A_{3}(\beta_{i},T_{i})=-6\beta_{i}^{4}(1+T_{i}^{2})(5+T_{i}^{2}),\\
    \\
    A_{i3}\equiv A_{4}(\beta_{i},T_{i})=\beta_{i}^{4}(1+T_{i}^{2})^{2},\\
    \\
    A_{i4}\equiv A_{5}(\beta_{i},T_{i})=-16\beta_{i}^{3}T_{i}^{3},\\
    \\
    A_{i5}\equiv A_{6}(\beta_{i},T_{i})=-96\beta_{i}^{2}(1+T_{i}^{2}),\\
    \\
    A_{i6}\equiv A_{6}(\beta_{i},T_{i})=48\beta_{i}^{2}(2+T_{i}^{2}),\\
    \\
    A_{i7}\equiv A_{8}(\beta_{i},T_{i})=64,\\
    \\
    B_{i}\equiv -A_{1}(\beta_{i},T_{i})=-\beta_{i}^{6}(1+T_{i}^{2})^{2}.
    \end{array}
\end{equation}
Когато седемте точки от контура на сянката са така подбрани, че $\det A_{ij}\neq 0$, то съответните им седем решения $q_{j}(\beta_{i},T_{i})$ на (\ref{sysqjLin}), са:
\begin{equation}\label{qKerr}
    \begin{array}{l}
    \mathbf{q}=\mathbf{A^{-1}}\centerdot\mathbf{B}\equiv\mathbf{v},\\
    \\
    \mathbf{q}\equiv\left(q_{1},q_{2},q_{3},q_{4},q_{5},q_{6},q_{7}\right),\\
    \\
    \mathbf{B^{T}}\equiv\left(B_{1},B_{2},B_{3},B_{4},B_{5},B_{6},B_{7}\right).
    \end{array}
\end{equation}
Съответната на (\ref{sysasmin}), разрешима относно масата $M$, спин параметъра $a$ на Керовската черна дупка и синуса от ъгъла на инклинация $s$, система полиномиални уравнения, би могла да е тази за $q_{1}$, $q_{2}$ и $q_{4}$, което означава, че $\left(Q_{1},Q_{2},Q_{3}\right)=\left(q_{1},q_{2},q_{4}\right)$. Съгласно (\ref{qj}) имаме:
\begin{equation}\label{asKerr}
    \begin{array}{l}
    M=\sqrt{v_{2}},\\
    \\
    a=\dfrac{\sqrt{v_{2}^{3}-v_{1}^{\ }v_{4}^{2}}}{v_{2}}\ ,\\
    \\
    \sin\theta_{0}=\pm\dfrac{v_{4}}{\sqrt{v_{2}^{3}-v_{1}^{\ }v_{4}^{2}}}\ .
    \end{array}
\end{equation}
Тъй като така определените стойности на $M$ и $a$ зависят от мащаба в който е начертан контурът на сянката на нашата черна дупка ($M=\widetilde{M}M_{u}$, $a=\widetilde{a}M_{u}$), то ще е добре да изразим чрез $v_{1}$, $v_{2}$ и $v_{4}$ тяхното отношение $a_{*}\equiv a/M$. Имаме:
\begin{equation}\label{aM}
  a_{*}\equiv\dfrac{a}{M}=\dfrac{\sqrt{v_{2}^{3}-v_{1}^{\ }v_{4}^{2}}}{v_{2}\sqrt{v_{2}}}\ .
\end{equation}
Другите четири решения за $q_{j}$ на линейната система (\ref{sysqjLinKerr}) задават връзки между различните $\beta_{i}$ и $T_{i}$ на избраните седем точки от контура на сянката, които са конкретни съотношения, отнасящи се за сянката на Керовската черна дупка. С други думи, за всеки седем $\beta_{i}$ ($\beta_{i}\neq 0$) и $T_{i}\equiv\tan\varsigma_{i}$ от контура на сянката на Керовската черна дупка, са валидни следните четири свързващи ги съотношения:
\begin{equation}\label{q37q12Kerr}
    \begin{array}{l}
    v_{3}=\dfrac{\left(v_{1}+1\right)^{2}v_{4}^{2}}{v_{2}^{2}}+\dfrac{4v_{2}^{4}}{v_{4}^{2}}-2\left(16v_{1}+1\right)v_{2},\\
    \\
    v_{5}=\dfrac{v_{1}\left(v_{1}+1\right)v_{4}^{2}}{v_{2}}+\dfrac{v_{2}^{5}}{v_{4}^{2}}-\left(2v_{1}+1\right)v_{2}^{2}\geq 0,\\
    \\
    v_{6}=v_{2}^{2}-\dfrac{\left(v_{1}+1\right)v_{4}^{2}}{v_{2}}\geq 0,\\
    \\
    v_{7}=\dfrac{1}{v_{4}^{2}}\left[v_{2}^{2}-\dfrac{\left(v_{1}+1\right)v_{4}^{2}}{v_{2}}\right]^{3}\geq 0.
    \end{array}
\end{equation}
При това $v_{i}=0$ за всяко $i>4$, когато ъгълът на инклинация $\theta_{0}=\pi/2$.

Когато черната дупка на Kerr не е гола сингулярност ($a\leq M$), координатата $\beta$ достига максимум $b\equiv\beta_{max}$ при който $\beta'=0$ и съответно $\tan\varsigma\equiv T=0$. Можем да наречем $b$ "вертикален полудиаметър"\ на регулярната черна дупка. Друга характерна стойност на $\beta$ от контура на сянката на регулярната черна дупка на Kerr, е $b_{0}\equiv\beta(r_{0})$ за която $\alpha\equiv\alpha(r_{0})=0$. Удобно е да изразим тeзи две отсечки от контура на сянката на регулярната черна дупка на Kerr чрез $M$, $a$ и $s$. Зависимостта на $b$ от параметрите $M$, $a$ и $s$ се получава непосредствено от (\ref{solP}) при $T=0$. Имаме следното кубично уравнение за $b^{2}$:
\begin{equation}\label{eqb}
  b_{*}^{4}\left(b_{*}^{2}-27\right)-a_{*}^{2}b_{*}^{4}\mathfrak{c}^{2}\left(b_{*}^{2}-30\right)+
  a_{*}^{4}b_{*}^{2}c^{4}\left(b_{*}^{2}-96\right)+64a_{*}^{6}c^{6}=0,
\end{equation}
където $a_{*}\equiv a/M$, $b_{*}\equiv b/M$ и $c\equiv\cos\theta_{0}$. Намирането на връзката между $b_{0}$, $M$, $a$ и $s$ е важно, защото ако вече сме получили стойностите на параметрите $M$, $a$ и $s$, то с помощта на $b_{0}$ можем да установим положението на "вертикалната"\ ос $\beta$, което ще позволи да определяме координатите на точките от контура на сянката в "хоризонтално"\ направление. Нека $T_{0}\equiv T(r_{0})$. Тогава съгласно (\ref{PolysysalphabetaT}):
\begin{equation}\label{Polysysalphab0T0}
    \begin{array}{ll}
    P_{\alpha}\left(r_{0}\right)\equiv -r_{0}^2(r_{0}-3M)-a^2(r_{0}+M)=0,\\
    \\
    P_{\beta}\left(r_{0}\right)\equiv(b_{0}^{2}+a^{2}s^{2})(r_{0}-M)^{2}-2\left[r_{0}^{2}(r_{0}^2-3M^2)+a^2(r_{0}^2+M^2)\right]=0,\\
    \\
    P_{T}\left(r_{0}\right)\equiv b_{0} T_{0}(r_{0}-M)-as(r_{0}+M)=0.
    \end{array}
\end{equation}
За да намерим връзката между $b_{0}$, $M$, $a$ и $s$, следва в (\ref{Polysysalphab0T0}) да елиминираме $r_{0}$ и $T_{0}$. Получаваме:
\begin{equation}\label{solb0}
    \begin{array}{ll}
    -a_{*}^{8}s^{4}\left(1-s^{2}\right)-
    a_{*}^{6}\left[s^{6}+30s^{4}-96s^{2}+64+b_{0*}^{2}s^{2}\left(2-3s^{2}\right)\right]+\\
    \\
    +a_{*}^{4}\left[27s^{4}-3b_{0*}^{2}\left(s^{4}+20s^{2}-32\right)+b_{0*}^{4}\left(3s^{2}-1\right)\right]+\\
    \\
    a_{*}^{2}b_{0*}^{2}\left[54s^{2}-3b_{0*}^{2}\left(s^{2}+10\right)+b_{0*}^{4}\right]-b_{0*}^{4}\left(b_{0*}^{2}-27\right)=0,
    \end{array}
\end{equation}
където $b_{0*}\equiv b_{0}/M$. По същият начин от (\ref{Polysysalphab0T0}) като елиминираме $r_{0}$ и $b_{0}$, ще намерим как $T_{0}$ зависи от $M$, $a$ и $s$. Резултатът е:
\begin{equation}\label{solT0}
    \begin{array}{ll}
    a_{*}^{2}\left[s^{6}\left(1+T_{0}^{2}\right)^{2}\left(4+T_{0}^{2}\right)+
    s^{4}T_{0}^{2}\left(57T_{0}^{4}-78T_{0}^{2}+9\right)+96s^{2}T_{0}^{4}\left(1-T_{0}^{2}\right)+\right.\\
    \\
    \left.+64T_{0}^{6}\right]-a_{*}^{4}T_{0}^{2}\left[2s^{6}\left(T_{0}^{4}+2T_{0}^{2}-7\right)+
    s^{4}\left(29T_{0}^{4}-38T_{0}^{2}+13\right)+\right.\\
    \\
    \left.+32s^{2}T_{0}^{2}\left(1-3T_{0}^{2}\right)+64T_{0}^{4}\right]-
    s^{4}T_{0}^{2}\left(1+T_{0}^{2}\right)^{2}\left[a_{*}^{6}\left(1-s^{2}\right)-27\right]=0.
    \end{array}
\end{equation}

\section*{\begin{center}Acknowledgments\end{center}}

The author acknowledges support by the Bulgarian NSF Grant No. KP-06-N28/7.

\end{document}